# Timing properties of XB 1254–690

Sudip Bhattacharyya[1,2]⋆
[1] *CRESST and X-ray Astrophysics Laboratory, NASA/GSFC, Greenbelt, MD 20771, USA*
[2] *Department of Astronomy, University of Maryland, College Park, MD 20742, USA*

**ABSTRACT**

We analyze archival Rossi X–ray Timing Explorer (*RXTE*) Proportional Counter Array (PCA) data of the low mass X–ray binary (LMXB) system XB 1254–690. We calculate colour-colour diagram, hardness-intensity diagram and power spectra of this source, associate its broad low-frequency timing features with the portions of the colour-colour diagram, and establish that XB 1254–690 is an atoll source. This will be important to correlate the timing and spectral features of this dipping LMXB with its various states, which may be useful to understand LMXBs in general. We find the source always in the high intensity banana state, which may explain why $\sim 1$ Hz "dipper QPOs" have never been observed from XB 1254–690. We also report a suggestive evidence of millisecond period brightness oscillations at the frequency $\sim 95$ Hz during a thermonuclear X–ray burst for the first time from this source.

**Key words:** methods: data analysis — stars: neutron — techniques: miscellaneous — X-rays: binaries — X-rays: bursts — X-rays: individual (XB 1254–690)

## 1 INTRODUCTION

XB 1254–690 is a persistent low mass X–ray binary (LMXB) system (Griffiths et al. 1978), that exhibits energy dependent intensity dips (Courvoisier et al. 1986) with the binary orbital period ($\sim 3.88$ hr). This modulation of soft X-ray intensity is believed to be caused by structures above the accretion disc (White & Swank 1982). This is possible only if the dippers are high inclination systems, so that the line of sight passes through these structures. Therefore, dippers provide an opportunity to constrain the properties of the upper layers of accretion discs (and the photoionized plasma above them; Jimenez-Garate et al. 2003) in LMXBs. However, in order to understand the nature of these X–ray emitting and absorbing components, it is essential to identify various states of the source, and to correlate the observed timing and spectral features with these states. But, no detailed timing properties of XB 1254–690 have previously been reported. Moreover, the nature of this source is not yet established (e.g., Z, atoll, etc.; see, for example, van der Klis 2004; Kuulkers et al. 1997; Wijnands et al. 1998; Homan et al. 2002; Agrawal & Bhattacharyya 2003). Here we note that, although Gladstone, Done, Gierliński (2006) have recently included XB 1254–690 in the list of atoll sources, these authors have not given a discussion to justify this, have not shown a separate colour-colour diagram (van der Klis 2004; Belloni & Hasinger 1990; Olive, Barret, Gierliński 2003; Jonker et al. 2000) of this source, and have not done timing analysis using the *RXTE* PCA data of XB 1254–690 in order to find the nature of this source.

XB 1254–690 also exhibits type-I X–ray bursts (Mason et al. 1980; Courvoisier et al. 1986). These bursts are produced by thermonuclear burning of matter accumulated on the surfaces of accreting neutron stars (Woosley & Taam 1976; Lamb & Lamb 1978), which shows that XB 1254–690 contains a neutron star. During many such bursts from some other LMXBs, the combination of rapid stellar rotation and an asymmetric brightness pattern on the neutron star surface gives rise to observable millisecond period brightness oscillations (Strohmayer & Bildsten 2006). The frequency of these oscillations gives the neutron star spin frequency (Chakrabarty et al. 2003). Moreover, this timing feature may be useful to constrain stellar structure parameters (Bhattacharyya et al. 2005; Nath, Strohmayer, Swank 2002), and to understand thermonuclear flame spreading on neutron stars (Bhattacharyya & Strohmayer 2005; 2006a; 2006b; 2006c). However, until now, burst oscillations from XB 1254–690 have never been reported, and the spin frequency of the neutron star is not known. Therefore, the detection of these oscillations from this source will be very useful.

In this paper, we perform timing and spectral analysis of the *RXTE* PCA data of XB 1254–690, calculate the colour-colour diagram (van der Klis 2004), correlate broad timing features with different portions of it, and establish that XB 1254–690 is an atoll source. Moreover, we find a suggestive evidence of oscillations from one type-I (thermonuclear) X-ray burst. In § 2, we describe our data analysis procedure

⋆ E-mail: sudip@milkyway.gsfc.nasa.gov

and results, and in § 3, we discuss the implications of these results.

## 2 DATA ANALYSIS AND RESULTS

The LMXB system XB 1254–690 was observed with *RXTE* in the years (proposal numbers) 1996 (P10068), 1997 (P20066), 2001 (P60044) and 2004 (P90036). In 2001, the source was observed both in May, and in December. We have used these proportional counter array (PCA) instrument data to calculate the colour-colour diagram (CD; Fig. 1), and hardness-intensity diagram (HID; Fig. 1), and to perform timing studies. For the first two, we have always used PCA Standard 2 mode data, while for the third one we have used PCA Good Xenon data (except for 1996 observations, for which we have used PCA Event mode data with about 16 $\mu$s resolution). The PCUs 0, 2 and 3 were reliably on during these observations, and we have used the data from these PCUs (and only from the top Xenon layers to increase signal-to-noise) to calculate the CD and HID. We have defined the soft and hard colours for the CD (the intensity for the HID) as the ratio of the background-subtracted counts (background-subtracted count) in the channel and energy ranges mentioned in Table 1. These definitions for various observations are very close to each other, and also very similar to those used by Muno, Remillard & Chakrabarty (2002; hereafter MRC). This will facilitate the comparison of our results with those of these authors. However, the gain settings of PCUs change from time to time, and small gain shift happens even in the same gain epoch. As a result, artificial relative differences among CD (and HID) tracks of various observations may occur. We have corrected this effect by using the PCA data of Crab pulsar/nebula, as the Crab colours can be supposed to be constant (Homan et al. 2002; van Straaten et al. 2000; Altamirano et al. 2005). We have scaled the colours of the XB 1254–690 observation sets (except the one during May, 2001) mentioned in Table 1 to correctly compare them with those of the May, 2001 observations. We have utilized the Crab data at the closest time (but in the same gain epoch) to each of the five observation sets to do this (see, for example, Homan et al. 2002).

In order to explore the nature of XB 1254–690, first we have calculated a typical X–ray flux from this source by fitting spectra with a model. The $2-20$ keV average flux is $\approx 10^{-9}$ ergs cm$^{-2}$ s$^{-1}$. Considering an estimated bolometric correction factor $\sim 2$, and assuming (1) the source distance in the range $8-15$ kpc (Motch et al. 1987), (2) the neutron star mass in the range $M_{\rm x} = 1.4 - 2.0 M_\odot$, and (3) solar hydrogenic abundance, we have found the source luminosity in the range $\approx 0.05 L_{\rm Edd} - 0.25 L_{\rm Edd}$, where $L_{\rm Edd}$ is the Eddington luminosity.

We have, then, compared our CD (Fig. 1) with that of MRC. Fig. 1 of MRC shows that the flaring branches (FBs) of Z sources have hard colours less than 0.3 (except for GX 17+2). As the hard colour of XB 1254–690 is always greater than 0.3, this indicates that if this source is a Z source, it was not in an FB state. Besides, during these observations the source showed frequent flares, which are not usual for the other two states (normal branch and horizontal branch) of Z sources. Therefore, it is not likely that XB 1254–690 is a Z source. Moreover, the typical luminosities of Z sources are close to the Eddington luminosity (van der Klis 2004), and hence much higher than the luminosity of XB 1254–690. Therefore, we conclude that XB 1254–690 is not a Z source.

Fig. 1 of MRC shows that the hard colours of atoll sources are always greater than 0.3 (except for GX 13+1). This, and especially the observed hard colours of the banana branches of atoll sources (see MRC) strongly indicate that XB 1254–690 is an atoll source, and it was in the banana branch during all the observations (see Fig. 1). The following findings also suggest that XB 1254–690 was in banana branch during these observations: (1) the source covered a large part of its CD track in a few hours, and (2) the source hardened as its intensity increased (i.e., possibly went into upper banana (UB) state from lower banana (LB) state; van der Klis 2004).

To confirm the identification of the banana state of XB 1254–690, we have computed and fitted the power spectra of the source during low intensity and high intensity for each of 2001 and 2004 data (Table 2). The chosen 2001 data sets are from the same ObsID, which shows that the source intensity can significantly change with a timescale of hours. Both the power spectra from these low intensity and high intensity data sets can be well fitted with a 'constant+powerlaw' model (no. 1 & 2 of Table 2). The 'constant' describes the Poisson noise level, and the 'powerlaw' describes the very low frequency noise (VLFN). The detection of the latter noise strongly indicates that the source was in banana state during the 2001 observations (van der Klis 2004). Moreover, the strength of the VLFN increases with the intensity and the hardness of the source. This, and the disappearance of the VLFN during the low intensity state of the 2004 data (Table 2 and Fig. 2) are also consistent with banana state properties of atoll sources (van der Klis 2004). We also note that the power spectrum of the low intensity 2004 data can be well fitted with a 'constant+Lorentzian' (Fig. 2). The 'Lorentzian' indicates a broad hump, or alternatively a band-limited noise (BLN; van der Klis 2004).

We have detected five type-I (thermonuclear) X–ray bursts from XB 1254–690: one from 1996 data and four from 2001 data. All of them are non-photospheric-radius-expansion bursts with the peak count rates in the range $\sim 500 - 800$ counts s$^{-1}$ PCU$^{-1}$ (for all the detector channels). None of the 2001 bursts shows significant millisecond period brightness oscillations. However, we have found suggestive evidence of burst oscillations during the rising portion of the 1996 burst. In the inset panel of Fig. 3, we show the first power spectrum that we have calculated using 16 $\mu$s event mode data and a 1 s time interval (i.e., 1 Hz frequency resolution). This interval starts at the time when the burst count rate starts rising sharply. We have found a candidate peak at $\sim 95$ Hz with the power of $\approx 24.3$. The probability of obtaining a power this high in a single trial from the expected $\chi^2$ noise distribution (2 dof) is $\approx 5.29 \times 10^{-6}$. Multiplying this with the number of trials ($N_{\rm trial}$) would give the significance of this peak. The Nyquist frequency = 2048 Hz of our power spectrum, and five observed bursts from XB 1254–690 give $N_{\rm trial} = 10240$ for 1 Hz frequency resolution. This implies that the oscillations (at $\sim 95$ Hz) are detected with $\approx 95\%$ confidence. Although, the calculation of oscillation significance separately for each burst (as the presence of oscillations depends on burst properties and accretion history; Galloway et al. 2006) would imply a detection from the 1996 burst with 98.9% confidence ($N_{\rm trial} = 2048$), the more conservative value (95%) should be acceptable, because of the lack of our complete understanding of burst oscillations.

We have also fitted the phase-folded lightcurve of 1 s of data (the same data used to calculate the power spectrum of Fig. 3) with a combined model of a sinusoid and a constant ($\chi^2/{\rm dof} = 3.2/13$), and found that the oscillation rms amplitude is $0.31 \pm 0.07$. These values of fractional rms amplitude are consistent with that expected for a hot spot on a spinning neutron star surface (Miller & Lamb 1998). These values are also consistent with the rms amplitude value ($= 0.27$) obtained from the formula $(P_{\rm n}/I)^{1/2}(I/(I-B))$ (Muno, Özel, Chakrabarty 2003). Here $P_{\rm n}$ is the power at the nth bin of the Fourier spectrum, $I$ is the total number of counts in the profile, and $B$ is the estimated number of background counts in the profile. We have not found any significant burst oscillation frequency evolution, as were seen from some other bursting sources (e.g., Strohmayer et al. 1998; Bhattacharyya & Strohmayer 2006c; Bhattacharyya et al. 2006a).

We have searched for kHz QPOs in the whole data set, and found a tentative indication of ≈1000 Hz QPOs in the ObsID 20066-01-01-04. The significance of these plausible kHz QPOs is $\approx 2.66\sigma$. The high centroid frequency, the low $Q$-value ($\sim 2.44$), and the inferred fractional rms amplitude ($\approx 0.12$) are consistent with those of upper frequency kHz QPOs observed from other sources (see, for example, Barret, Olive, Miller, 2005). This plausible timing feature appeared in the relatively lower portion (that is lower soft colour and lower hard colour) of the banana branch. This is consistent with what has been seen in other atoll sources, because kHz QPOs are normally found in the "lower left banana" portion of the CD (van der Klis 2004). If confirmed, this timing feature will be the first kHz QPOs observed from this source. However, more significant detection of this feature is essential, as this plausible kHz QPOs appeared only in a small ($\sim 400$ s) data set.

## 3 DISCUSSION

In this paper, we have estimated the luminosity range (in the unit of Eddington luminosity) of XB 1254–690, calculated the colour-colour diagram, hardness-intensity diagram and persistent emission power spectra of this source, and established that this source is an atoll source. This is an important first step to identify various states of this source. As we have mentioned in § 1, XB 1254–690 is a dipping LMXB, and hence can be particularly useful to understand the properties of the upper layers of accretion discs, and the photoionized plasma above them in LMXBs. Moreover, spectral lines are normally observed from the dippers, and X–ray absorption lines were actually detected from XB 1254–690 (Boirin et al. 2004). Therefore, this source can be very promising for understanding LMXBs, if its timing and spectral features can be correlated to its various states. In this paper, we have already associated some low-frequency broad timing features with the LB and UB states of XB 1254–690. These, as well as the weak indication of a kHz QPO (see § 2), will encourage more observations of this source. Moreover, dippers can exhibit $\sim 1$ Hz QPOs, and to the best of our knowledge, such QPOs have been detected from four dippers so far (Jonker, van der Klis, Wijnands 1999; Homan et al. 1999; Jonker et al. 2000; Bhattacharyya et al. 2006c). However, these QPOs are observed at the low intensity states, which might explain why they were never observed from XB 1254–690, as according to our analysis, this source was always in the high intensity banana state during *RXTE* observations. This may motivate *RXTE* observations of XB 1254–690 at low intensities.

We have found a suggestive evidence of burst oscillations from XB 1254–690, and if confirmed (by the detection of more bursts), these will provide the neutron star spin frequency. The plausible oscillations were observed from only one burst out of the five observed from this source, although the other properties of all these bursts seem similar. This is typical of other burst oscillation sources, and may be understood from the the recent work of Spitkovsky et al. (2002). According to these authors, the low-latitude ignition of bursts is much more likely, and oscillations do not occur for such ignitions. This supports the calculation of oscillation significance separately for each burst, although a better understanding of burst oscillations is required to strongly justify such calculation (see § 2). Finally, if the 95 Hz burst oscillations from XB 1254–690 are confirmed, the neutron star with low spin frequency (inferred from the oscillations) will be ideal to search for surface atomic spectral absorption lines (see Cottam, Paerels, Méndez 2002; Bhattacharyya, Miller, Lamb 2006b), which may have sufficient depths for detection.

## ACKNOWLEDGMENTS

The author thanks Tod E. Strohmayer for detailed discussion, and an anonymous referee for helpful suggestions. This work was supported in part by NASA Guest Investigator grants.

4     *Sudip Bhattacharyya*

**Table 1.** Channel and energy boundaries of the soft colour, hard colour, and intensity used in Figure 1.

| Year | Channels[a] (Soft colour) | keV (Soft colour) | Channels[a] (Hard colour) | keV (Hard colour) | Channels[a] (Intensity) | keV (Intensity) |
|---|---|---|---|---|---|---|
| 1996 | 10–13/6–9 | 3.71–5.15/2.29–3.71 | 24–49/14–23 | 8.74–18.23/5.15–8.74 | 6–49 | 2.29–18.23 |
| 1997 | 10–13/6–9 | 3.68–5.11/2.26–3.68 | 24–49/14–23 | 8.70–18.18/5.11–8.70 | 6–49 | 2.26–18.18 |
| 2001 (May) | 9–12/6–8 | 3.58–5.28/2.30–3.58 | 21–42/13–20 | 8.69–18.16/5.28–8.69 | 6–42 | 2.30–18.16 |
| 2001 (Dec) | 9–12/6–8 | 3.55–5.25/2.29–3.55 | 21–42/13–20 | 8.65–18.09/5.25–8.65 | 6–42 | 2.29–18.09 |
| 2004 | 9–12/6–8 | 3.45–5.12/2.21–3.45 | 21–43/13–20 | 8.46–18.15/5.12–8.46 | 6–43 | 2.21–18.15 |

[a] The channel numbers refer to the absolute PCA channels (0–255).

**Table 2.** Best fit parameters (with $1\sigma$ error) for the low frequency (up to $\sim 100$ Hz) *RXTE* power spectra from XB 1254–690.

| No.[a] | PLN $\nu$[b] | PLN rms[c] (%) | $L_{f_0}$[d] (Hz) | $L_\lambda$[e] (Hz) | $L_{\rm rms}$[f] (%) | $\chi^2$/dof |
|---|---|---|---|---|---|---|
| 1 | $-1.17 \pm 0.08$ | $3.4 \pm 0.6$ | – | – | – | 65.88/80 |
| 2 | $-0.92 \pm 0.06$ | $4.5 \pm 0.5$ | – | – | – | 82.34/80 |
| 3 | – | – | $0.011 \pm 0.004$ | $0.018 \pm 0.007$ | $1.8 \pm 0.3$ | 22.67/18 |
| 4 | $-1.44 \pm 0.13$ | $2.8 \pm 1.0$ | – | – | – | 44.31/56 |

[a] 1: ObsID 60044-01-01-01 (lower intensity), 2: ObsID 60044-01-01-01 (higher intensity), 3: ObsID 90036-01-01-02 (low intensity), and 4: ObsID 90036-01-01-03 (high intensity). Power spectra are fitted either by constant+Lorentzian, or by constant+powerlaw in the energy range $\sim 2.6 - 30$ keV.
[b] Index of power law noise ($\propto f^\nu$; $f$ is frequency).
[c] RMS of power law; lower limit of integration is 0.004 Hz.
[d] Centroid frequency of Lorentzian ($\propto \lambda/[(f - f_0)^2 + (\lambda/2)^2]$).
[e] Full width at half maximum (FWHM) of Lorentzian.
[f] RMS of Lorentzian.

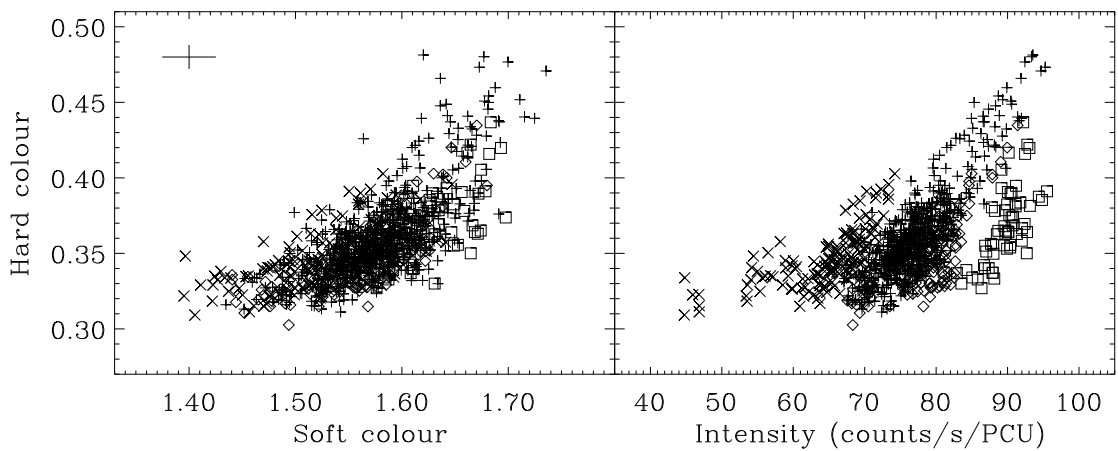

**Figure 1.** Colour-colour diagram (left panel) and hardness-intensity diagram (right panel) of XB 1254–690 using *RXTE* PCA data of the years 1996 (proposal number: 10068; *square* signs), 1997 (proposal number: 20066; *cross* signs), 2001 (proposal number: 60044; *plus* signs), and 2004 (proposal number: 90036; *diamond* signs). The definitions of the colours and the energy range of the intensity are given in Table 1. Here we use only the top Xenon layers of the PCUs 0, 2 and 3, and exclude the flaring and dipping data. Each point is for 128 s of data. Typical $1\sigma$ error bars for soft and hard colours are shown in the left panel. The ranges of soft colour, hard colour and intensity strongly indicate that XB 1254–690 is an atoll source.

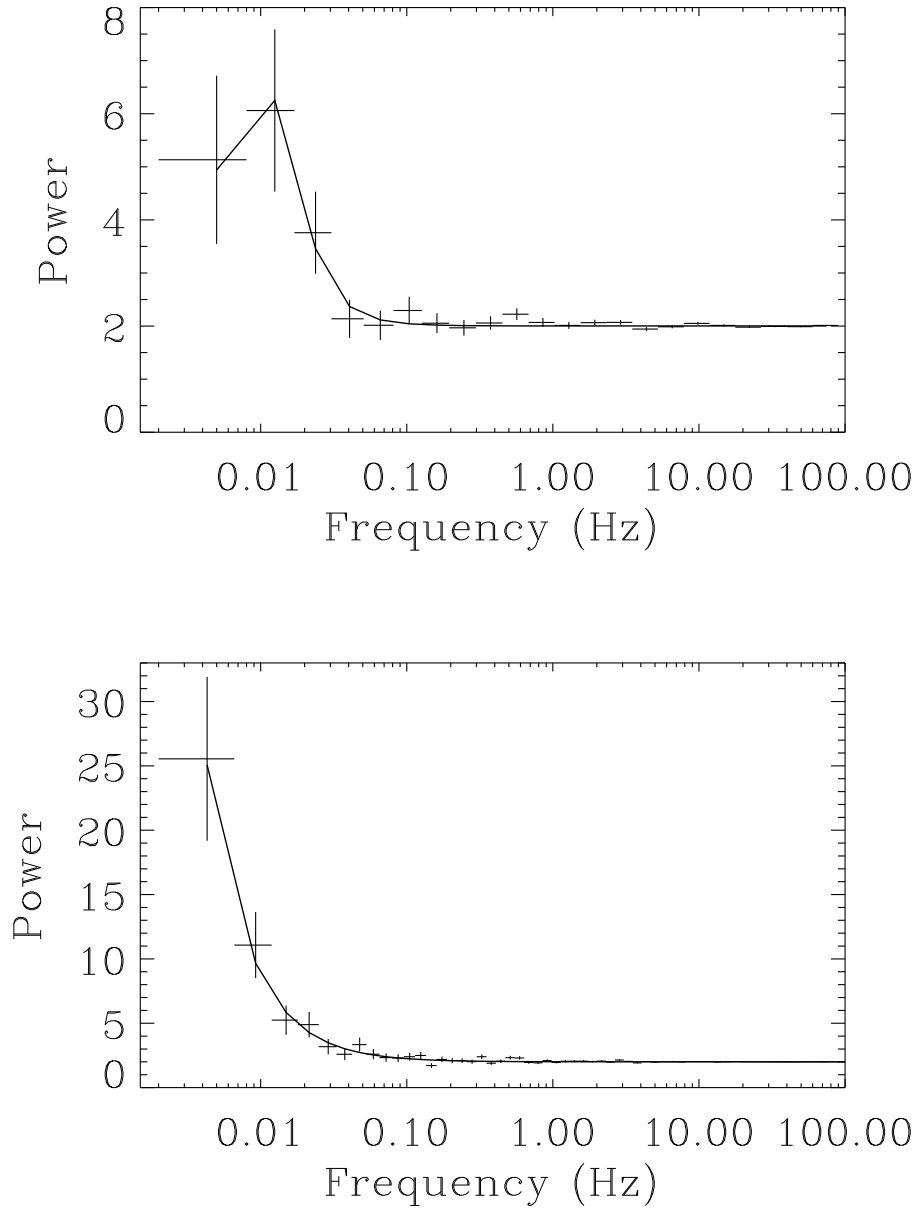

**Figure 2.** Low frequency power spectra (for the energy range $\sim 2.6 - 30$ keV) of XB 1254–690 using *RXTE* PCA data of 2004. The upper and lower panels are for the power spectrum numbers 3 (low intensity) and 4 (high intensity) of Table 2. Each panel shows the data points and the best fit model (solid line). The data of the upper panel are fitted well with 'constant+Lorentzian', while the best fit model for the lower panel data is 'constant+powerlaw'. The 'constant' describes the Poisson noise level, the 'Lorentzian' indicates a broad hump or a band-limited noise (BLN), and the 'powerlaw' describes the very low frequency noise (VLFN). These indicate that XB 1254–690 was in banana state during these observations. The horizontal lines around the data points give the frequency bin, and the corresponding vertical lines give the $1\sigma$ errors of powers.

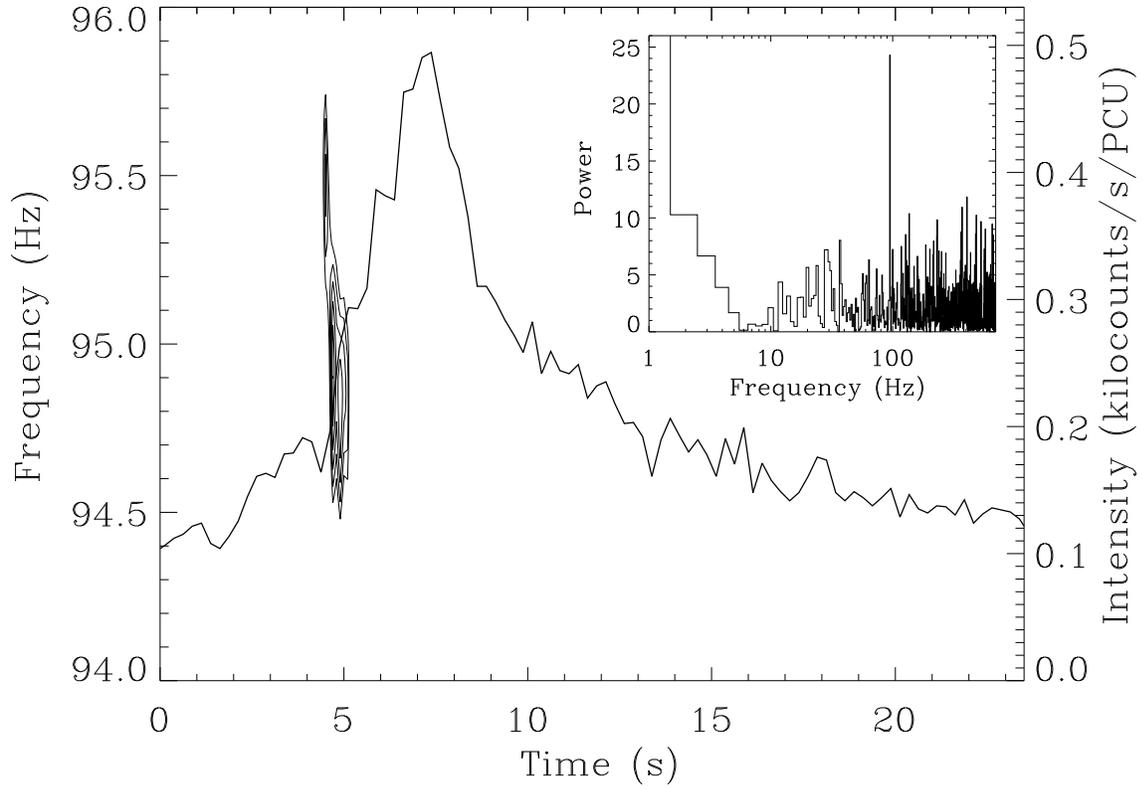

**Figure 3.** A thermonuclear X–ray burst from XB 1254–690 (1996 data). The main panel shows the PCA count rate profile and the power contours (of the plausible burst oscillations) using the dynamic power spectra (for 1.0 s duration at 0.1 s intervals; Strohmayer & Markwardt 1999). Contours at power levels of 19, 21, 23, 25, 27, 28.8 are given. The inset panel shows a power spectrum of 1 s interval (the first such interval after the burst count rate starts rising sharply) with 1 Hz frequency resolution. The peak near 95 Hz shows the plausible signal power.